\documentclass[runningheads]{llncs}
\usepackage[T1]{fontenc}
\usepackage{graphicx}
\usepackage{color}
\usepackage{booktabs}
\usepackage{graphicx}
\usepackage{xcolor}
\usepackage{multirow}
\usepackage{url}
\begin{document}
\title{QC-Stark: A Multi-Task Benchmark Revealing Capability Dissociations in LLMs Evaluated on Quantum Computing Tasks}
%
%
\author{Pranav Gupta\inst{1}\orcidID{0000-0002-1412-0885} }
\authorrunning{P. Gupta et al.}
%
\institute{Cisco, 2901 3rd Ave, Suite 600, Seattle, WA 98121, USA \\
\email{pranavpg@cisco.com}\\
}
\maketitle              
\begin{abstract}
We introduce QC-Stark, a benchmark for evaluating large language models (LLMs) on 11 quantum computing (QC) tasks, spanning circuit construction, debugging, compilation, error correction, and simulation.
Across 2,750 evaluations (10 models $\times$ 11 tasks $\times$ 5 difficulty levels $\times$ 5 seeds), we find that overall rankings mask substantial per-task variation. The Spearman correlation $\rho$ between overall and per-task rankings is statistically insignificant for 4 out of the 11 tasks included in this benchmark. A 2-parameter Item Response Theory (IRT) model validates measurement quality, and prompt sensitivity analysis confirms ranking robustness across prompt conditions.
All tasks are auto-verifiable via execution, thus not requiring any manual evaluation. We make the code and data publicly available on Huggingface.
\keywords{Large Language Models (LLMs)  \and Quantum Computing \and Item Response Theory.}
\end{abstract}
\section{Introduction}
While large language models (LLMs) are being increasingly applied to quantum computing (QC) tasks~\cite{cao2025selfdriving}, existing evaluations still focus on specific aspects, such as algorithm implementation~\cite{quantumkatas2026}, challenge-style programming~\cite{qcoder2025}, or conceptual question-answering~\cite{quantumaudit2026}.
There is a lack of benchmarks that span the complete spectrum of a quantum computing practitioner's day-to-day operations, ranging from circuit construction through compilation, debugging, error correction, and simulation. 

This gap is important because LLM capabilities cannot be expected to generalize across tasks, given the diversity in their architecture, algorithms and training/evaluation data.
A model that excels at oracle synthesis might score close to zero on hardware routing. In some cases, a model ranked first overall may score 0\% on a challenging task. An ideal benchmark should ideally also include features such as run-time procedural generation, in order to avoid dataset contamination or leakage. 

QC-Stark addresses these concerns by providing (1) 11 tasks covering the full QC workflow, (2) 5 difficulty levels per task, (3) seed-based procedural generation to resist against data contamination, and (4) automatic verification via code execution.

\section{Related Work} 
QC-Stark complements existing quantum computing benchmarks. Quantum-Audit~\cite{quantumaudit2026} evaluates conceptual knowledge via multiple-choice questions (best performance: 84\%), Qiskit QuantumKatas~\cite{quantumkatas2026} tests 350 educational exercises (best performance: 83\%), QCoder~\cite{qcoder2025} evaluates on contest problems with simulator feedback (best performance: 78\%), whereas QCircuitBench~\cite{qcircuitbench2024} provides large-scale algorithm design data. QC-Stark spans the full operational spectrum with systematic difficulty scaling in terms of 5 separate levels.

Outside quantum computing, scientific coding benchmarks like SciCode~\cite{scicode2024} (best performance: 4.6\%) and CMT-Benchmark~\cite{cmtbench2025} (best performance: 30\%) find frontier models far from research-level capability, thus implying the need for continuously developing new benchmarks which are fundamentally distinct from existing ones. 

\section{Benchmark Design}

\textbf{Tasks.} QC-Stark consists of 11 tasks organized by stages in a QC practitioner's workflow: Circuit Construction (State Preparation, Trotter Decomposition, Oracle Synthesis), Code Understanding (Debugging, Noise Discrimination, Reverse Engineering), Verification (Equivalence Checking), Compilation (Hardware Routing), Simulation (Noise Fidelity Estimation, Variational Quantum Eigensolver (VQE)), and Error Correction (Syndrome Decoding). Appendix~\ref{app:tasks} provides full descriptions for each task.

\textbf{Difficulty scaling.} Each task has 5 levels: L1 (\textsc{Textbook}), L2 (\textsc{Homework}), L3 (\textsc{Exam}), L4 (\textsc{Research}), L5 (\textsc{Open}), controlled via problem size, constraint complexity, and domain-specific parameters. Appendix~\ref{app:prompts} provides example prompt templates.

\textbf{Evaluation.} Each instance is deterministically generated from a (\texttt{task}, \texttt{level}, \texttt{seed}) tuple. Models receive a system prompt with Qiskit 2.x~\cite{qiskit241} API guidance and are asked to output a \texttt{solve()} function, which is run by a verifier to check for correctness. The output is matched against ground truth, in terms of aspects such as state fidelity, functional equivalence and correct identification. By principle, we do not require any human judgment.

\textbf{Models.} We evaluated 10 models: \texttt{o4-mini, Claude Sonnet 5, GPT-5.4, Gemma-4 31B, Claude Opus 4.1, GPT-4.1-mini, Gemini Flash Lite (Gemini FL), Gemini 3.5 Flash, Mistral Large 3,} and \texttt{LLaMA-3.3 70B}. These models were chosen based on our availability constraints and a goal of encompassing various model parameter sizes in our study. Note that we used Claude Code (Opus 4.6 and higher) for designing the benchmark, hence there could be some bias towards Claude models. In the future, this can be remediated by the use of a council of LLMs collectively designing the benchmark or by excluding those model families from the evaluation. We chose to keep Anthropic/Claude models for the sake of completion.

\section{Results}

\subsection{Overall Performance and Rank Inversions}

Table~\ref{tab:main} presents the scores from 2,750 evaluations (10 models $\times$ 11 tasks $\times$ 5 seeds $\times$ 5 difficulty levels). The best model (Claude Sonnet~5) achieves 0.662 overall, but its performance drops 59\% as we increase the difficulty level from L1 to L5. Task difficulty varies 5.9$\times$ (Equivalence: 0.78 vs.\ Debugging: 0.13).

\begin{table}[t]
\centering
\caption{Mean accuracy per model and task (all models at maximum supported token budgets). \textbf{Bold}: best per task. \color{gray}{Gray}: worst. \color{black} Last column: overall mean score $\pm$ standard\ deviation\ across seeds. Task codes: T1=State Preparation, T2=Trotter Decomposition, T3=Oracle Synthesis, T4=Debugging, T5=Noise Discrimination, T6=Reverse Engineering, T7=Equivalence, T8=Routing, T9=Noise Fidelity, T10=VQE, T11=Quantum Error Correction}
\label{tab:main}
\setlength{\tabcolsep}{2.5pt}
\footnotesize
\resizebox{\textwidth}{!}{
\begin{tabular}{lccccccccccc|c}
\toprule
Model & T1 & T2 & T3 & T4 & T5 & T6 & T7 & T8 & T9 & T10 & T11 & \textbf{Overall} \\
\midrule
Sonnet~5      & .68 & .24 & \textbf{1.00} & \textbf{.60} & \textbf{.64} & .64 & \textbf{1.00} & .20 & \textbf{1.00} & .60 & \textbf{.68} & \textbf{.662$\pm$.047} \\
Gemini~3.5F   & \textbf{.92} & \color{gray}{.00} & .96 & .48 & .60 & \textbf{.76} & .84 & .20 & \textbf{1.00} & .80 & .60 & .651$\pm$.059 \\
o4-mini       & .52 & \textbf{.64} & .84 & \color{gray}{.00} & .56 & .68 & .88 & .32 & \textbf{1.00} & .40 & \textbf{.68} & .593$\pm$.060 \\
GPT-5.4       & \color{gray}{.40} & .36 & .44 & .20 & .60 & .68 & .84 & \color{gray}{.04} & .80 & \textbf{.88} & .64 & .535$\pm$.083 \\
Gemma-4       & \color{gray}{.40} & .52 & .52 & \color{gray}{.00} & .52 & .48 & .76 & .28 & .72 & .56 & .64 & .491$\pm$.088 \\
GPT-4.1 mini      & \color{gray}{.40} & .20 & .24 & \color{gray}{.00} & .16 & .48 & .80 & .20 & .84 & .40 & .60 & .393$\pm$.034 \\
Opus~4.1      & \color{gray}{.40} & .24 & .48 & \color{gray}{.00} & \color{gray}{.00} & .32 & .88 & .16 & .76 & .56 & .40 & .382$\pm$.041 \\
Gemini~FL     & .44 & .04 & .36 & \color{gray}{.00} & .32 & .40 & \textbf{1.00} & \textbf{.44} & .48 & .48 & \color{gray}{.08} & .367$\pm$.067 \\
Mistral~L3    & \color{gray}{.40} & \color{gray}{.00} & .16 & \color{gray}{.00} & \color{gray}{.00} & .56 & .80 & .12 & \color{gray}{.04} & .40 & .40 & .262$\pm$.015 \\
LLaMA-70B     & \color{gray}{.40} & \color{gray}{.00} & \color{gray}{.00} & .04 & \color{gray}{.00} & \color{gray}{.04} & \color{gray}{.00} & .12 & \color{gray}{.00} & \color{gray}{.36} & .36 & \color{gray}{.120$\pm$.009} \\
\midrule
Task mean     & .50 & .22 & .50 & .13 & .34 & .50 & .78 & .21 & .66 & .54 & .51 & .445 \\
\bottomrule
\end{tabular}}
\end{table}

The overall ranking is \emph{not predictive} of per-task performance on 4 out of 11 tasks, where Spearman $\rho$ between the overall and individual task rankings is statistically non-significant. These 4 tasks were Routing ($\rho = +0.27$, $p = 0.45$), Trotterization ($\rho = +0.49$, $p = 0.15$), Debugging ($\rho = +0.54$, $p = 0.11$), and Equivalence ($\rho = +0.45$, $p = 0.19$). These inversions demonstrate that a single aggregate score is insufficient in guiding model selection for quantum computing tasks, and there is still progress to made for creating LLMs that outperform their predecessors on all kinds of benchmarks. 

\subsection{The ``Debugging'' Conundrum}

Despite all models being run at their maximum supported token budgets, 6 of 10 models score 0.00 on debugging. Claude Sonnet~5 leads at 60\% (15/25), followed by Gemini~3.5 Flash at 48\% (12/25 correct), GPT-5.4 at 20\% (5/25 correct), and LLaMA-70B at 4\% (1/25 correct). Dedicated reasoning models (o4-mini) score 0\% even at token budgets of 65{,}536 tokens. 
\subsection{Routing: A Universal Bottleneck}

Hardware routing (T8) has a mean accuracy of 0.21 across all models, with Gemini Flash Lite leading at a mean accuracy of 0.44 (driven by strong L1/L2 performance but weaker performance in more difficult problems in L3 and beyond). The Spearman correlation between overall and routing rankings is non-significant ($\rho = +0.27$, $p = 0.45$): the 8th-ranked model leads, while 4th-ranked GPT-5.4 scores merely 0.04 (1 is the maximum score in all these evaluations). Many evaluations produce circuits not functionally equivalent to input. This shows that models find it difficult to maintain circuit semantics while inserting SWAP gates for hardware connectivity.

\subsection{Measurement Validation}

\textbf{IRT analysis.} Our Item Response Theory (IRT) analysis follows ATLAS~\cite{helm2022}. Equally weighted averages over all tasks to determine the final rankings can thus be misleading. A 2-parameter logistic (2PL) IRT model~\cite{baker2004} fitted via Joint Maximum Likelihood Estimation (JMLE) yields marginal reliability $= 0.985$, with ability estimates spanning 3.6 standard units (Table~\ref{tab:irt_combined}). Of 275 items (11 tasks 
$\times$ 5 seeds $\times$ 5 difficulty levels), 216 (78.5\%) achieve discrimination $a > 1.0$, which provides evidence that our benchmark problems are appropriate in terms of difficulty. As expected, difficulty progresses monotonically from L1 ($b = -0.88$) to L5 ($b = +0.89$) (see Appendix~\ref{app:difficulty} for a per-level breakdown). The IRT ability ranking is nearly identical to the raw accuracy ordering ($\rho = 0.976$, $p < 0.001$). Claude Sonnet~5 leads in terms of both IRT ability ($\theta = 1.418$) and mean accuracy. The equation for the IRT model is given below.

\begin{equation}
\Pr(Y_{mjs}=1) \;=\; p_{mj} \;=\; \frac{1}{1 + e^{-a_j(\theta_m - b_j)}}
\label{eq:2pl}
\end{equation}

\begin{equation}
\log a_j \;\sim\; \mathcal{N}(0,\,\sigma^2), \qquad \sigma = 0.5, \qquad \lambda = 1/\sigma^2 = 4
\label{eq:prior}
\end{equation}

\begin{equation}
\ell(\theta,a,b) \;=\; \sum_{m=1}^{10}\sum_{j=1}^{55}
\Big[\, k_{mj}\log p_{mj} + (n_{mj}-k_{mj})\log(1-p_{mj}) \Big]
\label{eq:loglik}
\end{equation}

\begin{equation}
(\hat\theta,\hat a,\hat b) \;=\; \mathop{\rm arg\,max}\;
\Big\{\, \ell(\theta,a,b) \;-\; \frac{\lambda}{2}\textstyle\sum_{j=1}^{55}(\log a_j)^2 \Big\}
\label{eq:map}
\end{equation}

\noindent Here $m = 1,\dots,10$ indexes models, $j = 1,\dots,55$ indexes items
(one per task--level cell), and $s = 1,\dots,5$ indexes seeds.
$Y_{mjs} \in \{0,1\}$ is whether model $m$ solved seed $s$ of item $j$;
$p_{mj}$ is the probability of a correct response, shared by all five seeds of
a cell. $\theta_m$ is the ability of model $m$, standardised so that
$\theta$ has mean $0$ and unit standard deviation; $b_j$ is the difficulty of
item $j$ on that same scale, and $a_j > 0$ its discrimination, the slope of the
response curve in $\theta$. $k_{mj} = \sum_s Y_{mjs}$ is the number of seeds
solved and $n_{mj} = 5$ the number attempted. $\ell$ is the binomial
log-likelihood, $\sigma$ the prior standard deviation of $\log a_j$, and
$\lambda = 1/\sigma^2$ the equivalent penalty weight; hats denote estimates.

\begin{table}[t]
\centering
\caption{2PL IRT fit. Items are the 55 (task, level) cells, each administered to every model under 5 seeds; abilities are standardised to mean 0 and unit SD. 120 parameters (10 abilities, 55 difficulties, 55 discriminations) over 2750 responses. Discrimination carries a log-normal(0,\,0.5) prior, without which the 55 slopes are not identifiable from 10 models. In (b), $n$ is the number of identified items entering each mean: State Preparation L1 and L2 were passed by every model under every seed, leaving their difficulty unidentified, so they are excluded.}
\label{tab:irt_combined}
\footnotesize
\renewcommand{\arraystretch}{1.15}

{\smallskip (a) Model ability\par\smallskip}
\begin{tabular}{|c|l|r|r|r|}
\hline
\textbf{Rank} & \textbf{Model} & $\theta$ & \textbf{SE} & \textbf{95\% CI} \\
\hline\hline
1 & Gemini~3.5 & 1.122 & 0.088 & [0.949, 1.295] \\
\hline
2 & Sonnet~5 & 1.090 & 0.088 & [0.919, 1.262] \\
\hline
3 & o4-mini & 0.669 & 0.085 & [0.502, 0.837] \\
\hline
4 & GPT-5.4 & 0.432 & 0.087 & [0.262, 0.603] \\
\hline
5 & Gemma-4 & 0.251 & 0.089 & [0.077, 0.425] \\
\hline
6 & Opus~4.1 & $-$0.021 & 0.093 & [$-$0.203, 0.161] \\
\hline
7 & GPT-4.1m & $-$0.087 & 0.094 & [$-$0.272, 0.098] \\
\hline
8 & Gemini~FL & $-$0.141 & 0.096 & [$-$0.328, 0.046] \\
\hline
9 & Mistral L3 & $-$0.843 & 0.124 & [$-$1.086, $-$0.600] \\
\hline
10 & LLaMA-70B & $-$2.473 & 0.223 & [$-$2.909, $-$2.037] \\
\hline
\end{tabular}

\vspace{2.5ex}

{\smallskip (b) Task difficulty and discrimination\par\smallskip}
\begin{tabular}{|l|r|r|c|}
\hline
\textbf{Task} & \textbf{Mean difficulty} $\bar{b}$ & \textbf{Mean discrimination} $\bar{a}$ & $n$ \\
\hline\hline
Routing & 1.897 & 0.862 & 5 \\
\hline
Debugging & 1.711 & 2.267 & 5 \\
\hline
Trotter Decomposition & 1.581 & 0.987 & 5 \\
\hline
State Preparation & 1.116 & 2.841 & 3 \\
\hline
Noise Discrimination & 0.662 & 1.560 & 5 \\
\hline
Reverse Engineering & 0.107 & 1.349 & 5 \\
\hline
Oracle & 0.015 & 2.824 & 5 \\
\hline
QEC Decoding & $-$0.069 & 0.794 & 5 \\
\hline
Noise Fidelity & $-$0.325 & 2.526 & 5 \\
\hline
VQE & $-$0.917 & 1.243 & 5 \\
\hline
Equivalence & $-$1.272 & 1.597 & 5 \\
\hline
\end{tabular}
\end{table}

\textbf{Prompt sensitivity.} When we compared the canonical structured prompt against a second, minimal prompt across all 11 tasks and all 10 models (2,745 paired evaluations per condition), we found that model rankings are largely preserved ($\rho = 0.915$, Kendall $\tau = 0.778$). The top-4 ranked models stay in the top 4 across conditions, whereas the bottom-4 do not (GPT-4.1-mini falls from 6th to 8th under the minimal prompt, displacing Opus~4.1). Further details can be found in Appendix~\ref{app:prompt}.


\textbf{Error taxonomy and limitations.} All 2,750 evaluations were successful with 0 API errors after retries. Each model was run at its maximum supported token budget: 65{,}536 for o4-mini, GPT-5.4, Sonnet~5, Gemini~3.5 Flash, and Mistral~L3, 32{,}768 for GPT-4.1 mini, 32{,}000 for Opus~4.1, 8{,}192 for LLaMA-70B, and 4{,}096 for Gemma-4 and Gemini~FL. 11 records remained truncated at hard API caps that cannot be increased. 46 evaluations timed out during verification (we used a limit of 300 seconds), which were concentrated in VQE and Trotterization tasks that produce computationally expensive circuits. The benchmark requires Qiskit 2.x Python output, running the risk of conflating quantum domain knowledge with framework-specific API proficiency. A framework-agnostic format such as OpenQASM could potentially isolate reasoning but it changes task difficulty (e.g., state decomposition into elementary gates is substantially harder without high-level primitives). While the structured prompt's Qiskit 2.x guidance functions as lightweight context injection, retrieval augmented generation (RAG) with official documentation could further isolate reasoning ability from API memorization.

\vspace{-1mm}
\section{Conclusion}
\vspace{-1mm}

QC-Stark demonstrates that QC competence is not unidimensional: dedicated reasoning models o4-mini and o3-mini score 0\% on debugging even at 64K-token budgets (o3 scores 13\%). Surprisingly, Gemini Flash Lite (8th overall) leads routing at 44\%. The best accuracies per task display a huge range, from 13.2\% in Debugging to 78.0\% in Equivalence. Performance drops 59\% as we increase the difficulty level from L1 to L5. 

Prompt sensitivity analysis confirms ranking robustness ($\rho = 0.915$). IRT analysis validates the benchmark as a reliable instrument ($r = 0.985$) with monotonic difficulty progression, establishing that the observed dissociations reflect genuine structure in capability rather than just noise. Future work includes RLVR-based fine-tuning~\cite{rlvr2024}, framework-agnostic representations, and retrieval-augmented generation (RAG).

\section*{Acknowledgements}

Claude~\cite{claude2025} provided assistance for portions of benchmark infrastructure . All scientific claims and experimental design are solely the author's responsibility. The data for this benchmark is available at \url{https://huggingface.co/datasets/pranavgupta/qc-stark}.

\newpage
\appendix

\section{Prompt Sensitivity Details}
\label{app:prompt}

\begin{table}[h]
\centering
\caption{Prompt sensitivity: mean binarized-accuracy difference (structured $-$ minimal) by model and task, over all 11 tasks and all 10 models (2,745 paired evaluations per condition). The structured condition is the canonical system prompt used in the paper. 5 of the Gemma-4-31B State-Preparation L5 calls had to be excluded from this analysis: they could not be obtained in the minimal condition even after 6 retry rounds (persistent upstream 502/503 API errors). }
\label{tab:prompt_full}
\begin{tabular}{lr|lr}
\toprule
\textbf{Model} & \textbf{$\Delta$} & \textbf{Task} & \textbf{$\Delta$} \\
\midrule
o4-mini      & $+$0.113 & T11 (QEC Decoding)   & $+$0.244 \\
Gemma-4      & $+$0.107 & T9 (Noise Fidelity)    & $+$0.072 \\
GPT-4.1 mini    & $+$0.095 & T7 (Equivalence)   & $+$0.052 \\
Mistral L3   & $+$0.062 & T6 (Reverse Engineering)   & $+$0.048 \\
LLaMA-70B    & $+$0.058 & T8 (Routing)       & $+$0.044 \\
GPT-5.4      & $+$0.029 & T10 (VQE)          & $+$0.036 \\
Gemini FL    & $+$0.007 & T2 (Trotter Decomposition)       & $+$0.028 \\
Sonnet 5     & $+$0.000 & T3 (Oracle Synthesis)        & $+$0.004 \\
Opus 4.1     & $-$0.011 & T4 (Debugging)     & $-$0.016 \\
Gemini 3.5F  & $-$0.025 & T5 (Noise Discrimination) & $-$0.016 \\
             &          & T1 (State Preparation)    & $-$0.020 \\
\bottomrule
\end{tabular}
\end{table}

\noindent Spearman $\rho = 0.915$ (Kendall $\tau = 0.778$) between model rankings under the two conditions; overall accuracy rises 0.403 $\rightarrow$ 0.446 with the structured prompt. The top-4 grouping did not change across prompt types, but the bottom-4 ranks changed.

\section{Prompt Templates}
\label{app:prompts}

\subsection{System Prompts}

\noindent\textbf{Structured prompt:}

\begin{verbatim}
You are a quantum computing expert writing Python code.

Environment: Python 3.13, Qiskit 2.x, numpy, scipy.
Key Qiskit 2.x notes:
- QuantumCircuit.qasm() removed -> use qiskit.qasm2.dumps(circuit)
  / loads(qasm_str)
- execute() removed -> use Statevector or StatevectorSimulator
- qiskit.opflow removed -> use qiskit.quantum_info
  (Operator, Statevector)

Output requirements:
- Respond with ONLY executable Python code. No markdown fences, no
  explanations.
- Define a function called solve() that returns the answer.
\end{verbatim}

Note: The canonical string uses a Unicode right-arrow (U+2192) where \texttt{->} is shown above, and does not wrap the two long lines.

\noindent\textbf{Minimal prompt:}

\begin{verbatim}
You are a quantum computing expert. Respond with ONLY executable
Python code using Qiskit (version 2.x). No explanations, no markdown
fences, just the raw code. The code must define a function called
solve() that returns the answer. Do NOT use deprecated APIs like
.qasm() -- use qiskit.qasm2.dumps() if needed.
\end{verbatim}

\subsection{Task Prompt Examples}

\noindent\textbf{T1 --- State Preparation (L1, seed=1):}

\begin{verbatim}
Write a Qiskit function solve() that returns a QuantumCircuit on 2
qubits which prepares the following quantum state from |00>:

  |00>: 0.707106781186548 + 0.0i
  |01>: 0.0 + 0.0i
  |10>: 0.0 + 0.0i
  |11>: 0.707106781186548 + 0.0i

Requirements:
- You may use QuantumCircuit.initialize(statevector, qubits) or any
  standard gates.
- Ensure the statevector is normalized before passing to initialize().
- The circuit should achieve fidelity > 0.999 with the target state.
- Return ONLY the QuantumCircuit object from solve().
\end{verbatim}

\noindent\textbf{T4 --- Debugging (L2, seed=3):}

\begin{verbatim}
The following 3-qubit quantum circuit has exactly ONE bug (a wrong
gate, swapped qubits, or missing gate).

Buggy circuit (OpenQASM 2.0):
OPENQASM 2.0; include "qelib1.inc";
qreg q[3]; h q[0]; cx q[0],q[1]; cx q[0],q[2]; ...

The INTENDED unitary transformation maps basis states as follows:
  |000> -> (0.7071+0.0000i)|000> + (0.7071+0.0000i)|111>
  |001> -> ...

Identify the bug and write a Qiskit function solve() that returns
the CORRECTED QuantumCircuit.
\end{verbatim}

\section{Difficulty-Level Breakdown}
\label{app:difficulty}

\begin{table}[h]
\centering
\caption{Mean accuracy by difficulty level across all models and tasks.}
\label{tab:difficulty}
\begin{tabular}{lrrrrr}
\toprule
 & L1 & L2 & L3 & L4 & L5 \\
\midrule
Mean score & 0.655 & 0.553 & 0.416 & 0.336 & 0.267 \\
\bottomrule
\end{tabular}
\end{table}

\noindent Performance degrades monotonically with difficulty (59\% drop from L1 to L5). The steepest L1$\to$L5 drops occur on State Preparation ($1.00 \to 0.14$), VQE ($0.96 \to 0.12$), and Oracle Synthesis ($0.88 \to 0.24$). Debugging falls least in absolute terms ($0.16 \to 0.02$). Noise Fidelity is the only task that holds roughly flat across difficulty levels ($0.70 \to 0.66$).

\section{Task Descriptions}
\label{app:tasks}
\begin{table}[ht]
\centering
\caption{QC-Stark task descriptions with scoring criteria and difficulty scaling parameters.}
\label{tab:tasks_full}
\tiny
\begin{tabular}{p{3cm}p{3cm}p{3cm}p{4cm}}
\toprule
\textbf{Task} & \textbf{Description} & \textbf{Scoring} & \textbf{Difficulty Scaling} \\
\midrule
T1: State Preparation & Synthesize gate sequence producing target quantum state from $|0\rangle^n$ & State fidelity $> 0.999$ & L1--2: \texttt{initialize()} allowed; L3+: elementary gates only; number of qubits: 2--6 \\
\midrule
T2: Trotter Decomposition & Decompose $e^{-iHt}$ into Trotter product formula & Operator fidelity to exact evolution & Terms: 2--8; order: 1--4; time steps: 1--20 \\
\midrule
T3: Oracle Synthesis & Construct unitary oracle from Boolean specification & Functional equivalence to target & Input bits: 2--5; function complexity: linear to non-linear \\
\midrule
T4: Debugging & Locate and fix single injected fault in quantum circuit & Binary: correct circuit or not & Qubits: 3--7; fault types: gate substitution, qubit swap, gate deletion \\
\midrule
T5: Noise Discrimination & Distinguish noise-corrupted circuits from structurally-buggy ones & Correct identification & Channel types: depolarizing, amplitude damping; similarity level \\
\midrule
T6: Reverse Engineering & Identify algorithm from obfuscated circuit & Correct algorithm name & Algorithms: Bell, GHZ, QFT, Grover, VQE; obfuscation: transpiled to basis gates \\
\midrule
T7: Equivalence & Determine if two circuits implement same unitary & Binary correctness & Number of qubits: 2--5; circuit pairs: identical, permuted, distinct \\
\midrule
T8: Routing & Insert SWAPs for hardware connectivity & Functional equivalence on topology & Number of qubits: 5--20; topologies: linear, T, heavy-hex \\
\midrule
T9: Noise Fidelity & Construct noise channel matching error profile & Channel fidelity to target & Error types: depolarizing, amplitude damping, combined; number of qubits: 1--3 \\
\midrule
T10: VQE & Implement variational quantum eigensolver for ground-state energy estimation & Energy accuracy & Qubit count, ansatz depth, Hamiltonian complexity \\
\midrule
T11: QEC Decoding & Decode stabilizer syndrome to identify error & Correct error identification & Codes: repetition, Steane, surface; code distance: 3--7 \\
\bottomrule
\end{tabular}
\end{table}

\end{document}